\documentclass[aps,prl,twocolumn]{revtex4-1}

\usepackage{graphicx}
\usepackage{graphics}
\usepackage{dcolumn}
\usepackage{bm}
\usepackage{epstopdf}
\usepackage{mathrsfs}
\usepackage{amssymb}
\usepackage{amsmath}
\usepackage[pdftex]{hyperref}
\usepackage{natbib}
\usepackage{xspace}
\usepackage{slashed}
\usepackage{nicefrac}

\def\del{\partial}

\def\d{\text{d}}

\newcommand{\onehalf}{{\nicefrac{1}{2}}}

\newcommand{\qhat}{\ensuremath{{\hat q}}\xspace}
\newcommand{\pT}{\ensuremath{p_\perp}\xspace}

\newcommand{\beq}{\begin{eqnarray}}
\newcommand{\eeq}{\end{eqnarray}}
\newcommand{\be}{\begin{eqnarray*}}
\newcommand{\ee}{\end{eqnarray*}}

\begin{document}

\title{Jet (de)coherence in Pb-Pb collisions at the LHC}

\author{Yacine Mehtar-Tani}
\email{yacine.mehtar-tani@cea.fr}
\affiliation{%
Institut de Physique Th\'eorique,\\
CEA Saclay, F-91191 Gif-sur-Yvette, France}

\author{Konrad Tywoniuk}
\email{konrad@ecm.ub.edu}
\affiliation{%
Departament d'Estructura i Constituents de la Mat\`eria 
and Institut de Ci\`encies del Cosmos (IC-CUB),\\
Universitat de Barcelona, Mart\'i i Franqu\`es 1, 08028 Barcelona, Spain}

\date{\today}

\begin{abstract}
We study the modifications of jets created in heavy-ion collisions at LHC energies. 
The inherent hierarchy of scales governing the jet evolution allows to distinguish a leading jet structure, which interacts coherently with the medium as a single color charge, from softer sub-structures that will be sensitive to effects of color decoherence.
We argue how this separation comes about and show that this picture is consistent with experimental data on reconstructed jets at the LHC, providing a quantitative description simultaneously of the jet nuclear modification factor, the missing energy in di-jet events and the modification of the fragmentation functions. In particular, we demonstrate that 
effects due to color decoherence are manifest in the excess of soft particles measured in fragmentation functions in Pb-Pb compared to proton-proton collisions.

\end{abstract}


\maketitle

There is compelling evidence that a hot and dense quark-gluon plasma (QGP) is created in ultra-relativistic heavy ion collisions \cite{dEnterria:2009am}. Jets emerging from these collisions are unique probes of the underlying dynamics \cite{Mehtar-Tani:2013pia}. The features of a jet, such as its total energy $p_\perp$\footnote{Here, $p_\perp$ refers to the transverse momentum of the jet with respect to the beam axis and corresponds roughly to the energy of the jet since it is measured a mid-rapidity.}, derive from its constituents included within a reconstruction radius $\Theta_\text{jet}$. The distribution of hadrons in the jet is sensitive to interactions with the dynamical medium \cite{CasalderreySolana:2012ef}. Experimental results from Pb-Pb collisions at the LHC ($\sqrt{s_\text{NN}} = $ 2.76 TeV) have to date provided extensive data of the modifications of fully reconstructed jets with respect to their proton-proton baseline. On the one hand, a strong suppression of about $\sim$50 \% of the inclusive jet yield is observed in central collisions across a wide range in jet energy \cite{Aad:2012vca,*CMS:2012rba}. On the other hand, while the angular correlation of di-jet events is consistent with that in vacuum, the observed energy balance is strongly distorted and the missing energy of the leading jet is recovered only at very large angles with respect to the di-jet axis \cite{Chatrchyan:2011sx}. The jet fragmentation studies provide further detail, confirming to a large extent that the hard jet components escape the medium without large modifications while the soft components, typically occupying a broad angular range within the jet, are enhanced \cite{Spousta:2012rn,Chatrchyan:2012gw,CMS:2012vba,Chatrchyan:2013kwa}. 

In this letter, we present a comprehensive analysis of in-medium jet modifications based on perturbative QCD. We argue that a consistent picture that accommodates the three main trends observed in the data emerges. This comes about due to the large separation of the intrinsic jet scale $Q\equiv \Theta_\text{jet}p_\perp$ and the characteristic momentum scale of the medium $Q_s$, to be defined below, such that $Q \gg Q_s \gg \Lambda_\text{QCD}$, where $\Lambda_\text{QCD}$ is the non-perturbative scale of QCD. This scale separation is also intimately related to the interplay of two main mechanisms: induced independent gluon radiation off the coherent jet \cite{Baier:1994bd,Blaizot:2013hx}, which stands for the dominant medium effect, and corrections due to partial color decoherence of its constituents \cite{MehtarTani:2010ma,*MehtarTani:2011tz}. The former component is responsible for radiative energy loss and, in particular, for transporting this energy up to very large angles. The latter enhances the multiplicity of soft particles inside the jet, and hence is associated with radiation at relatively smaller angles. The basic parameters of the theory are related to the medium properties via the transport coefficient \qhat, which interrelates the momentum broadening and energy loss, and the in-medium mean free path $\lambda_\text{mfp}$. In addition, jet observables depend upon the geometry of the collision mainly through the mean path length $L$ for jet propagation. 

Relying on the collimation property of high-energy QCD jets \cite{Dokshitzer:1991wu,*Khoze:1996dn}, we will assume throughout this work that the medium does not resolve the leading sub-structures of the jet. In this case, the jet interacts coherently with the medium only via its total color charge and the ensuing medium-induced branchings shift the emitted energy up to very large angles with respect the the jet axis. This geometrical (angular) separation between vacuum-like and medium-induced branchings allows us to treat these two types of processes independent of each other and leads to Eq.~(\ref{eq:JetMedCalculus}), to be discussed in further detail below. Deviations from this simple picture are related to decoherence of vacuum-like radiation and corresponds to situations when probing sub-structures of the jet that are being resolved by the medium \cite{CasalderreySolana:2012ef}, see Eq.~(\ref{eq:JetMedTot}). We argue that this correction is essential for explaining key features of soft particles within the jet cone. The above considerations set out a strategy for enhancing the use jets to pin down medium effects.

The inclusive spectrum of reconstructed jets in Pb-Pb is suppressed compared to that in proton-proton \cite{Aad:2012vca,*CMS:2012rba}. At high $\pT$ this is caused by energy loss due to medium induced radiation \cite{Baier:1994bd}. Presently we will assume that these jets are mainly induced by primary quarks and parameterize the vacuum spectrum by a power law, $\d^2 \sigma^\text{jet}_\text{p-p} /\d^2 p_\perp \propto p_\perp^{-n}$, with the exponent $n\simeq 5.6$ extracted from experimental data \cite{Angerami:2012tr}. The nuclear modification factor is  defined as
\beq
\label{eq:QuenchingFactor}
R^\text{jet}_{AA}\equiv \frac{\d^2 N_\text{Pb-Pb}^\text{jet}(p_\perp)\big/ \d^2 p_\perp}{T_\text{AA}\,\d^2 \sigma_\text{p-p}^\text{jet}(p_\perp) \big/ \d^2p_\perp} \,,
\eeq
where $T_\text{AA}$ is the nuclear overlap function.
The inclusive spectrum of jets after passing the medium can be computed by convoluting the  jet cross-section in vacuum, proportional to the quark cross-section, with the distribution of quarks, $D^\text{med}_q$ after passing through the medium,
\beq
\frac{\d^2 N_\text{Pb-Pb}^\text{jet}(p_\perp) }{T_\text{AA}\, \d^2 p_\perp }\!\simeq \!\!\int_0^1 \!\frac{\d x}{x} D^\text{med}_{q} \!\left( x,\frac{p_\perp}{x},L \right) \! \frac{\d^2 \sigma^\text{jet}_\text{p-p}\left(\frac{p_\perp}{x} \right)}{ \d^2 p_\perp }\,,
\eeq
where $x$ is the fraction of the original quark energy carried by the quark after escaping the medium. For simplicity, the geometry of the collision is accounted for on average in terms of averaged values for $L$ and $\hat q$. Medium effects due to induced radiation encoded in the distribution of quarks are found by solving the following kinetic rate equation \cite{Jeon:2003gi,Blaizot:2013hx}
\begin{multline}
\label{eq:re_1}
\frac{\del}{\del L} D^\text{med}_i(x, p_\perp, L) = \int_0^1 \d z\, \mathcal{K}_{ij} \left(z, \frac{x}{z}p_\perp; L \right) \\ 
\times\left[D^\text{med}_j\left(\frac{x}{z}, p_\perp, L  \right) - z\,D^\text{med}_j(x,p_\perp,L) \right] \,,
\end{multline}
where the partonic distributions are $x\d N^\text{med}_i /\d x \equiv D^\text{med}_i(x,p_\perp, L)$ with $i = q,g$ \footnote{Presently we neglect the sub-leading feedback from gluon to quark conversion.}. The equation is governed by the branching rate, of parton $j$ into parton $i$,  per unit time, $\mathcal{K}_{ij}(z,p_\perp; t)$, which can be derived directly from the one-gluon emission spectrum \cite{Baier:1994bd}, 
\beq
\label{eq:BDMPSsimple}
\int_0^L \d t \, \mathcal{K}_{ij}(z,p_\perp; t) = \frac{\alpha_s}{2\pi} P_{ij}(z) \, \ln \left|  \cos \frac{(1+i) L} { 2\, t_\text{br} } \right| \,,
\eeq
where $\alpha_s$ is the strong coupling constant (in this work, $\alpha_s = 0.5$ \cite{Baier:2001yt}), $P_{ij}(z)$ are the (unregularised) Altarelli-Parisi splitting functions and $t_\text{br} \equiv \sqrt{ z(1-z)p_\perp/ \hat q _\text{eff} } $ is the branching time where $z$ is the fraction of the energy of parton $j$ carried by parton $i$. Finally, the effective transport coefficient probed in course of the branching is $\qhat_\text{eff} = \frac{1}{2}\big(1 + z^2 + [2C_2(j)/C_A - 1](1-z)^2 \big)\hat q$, where $C_2(j)$ is the color factor of the parton with label $j$, and \qhat is consistently referring to the quenching parameter in the adjoint representation. 

The form of the branching rate employed in this Letter, Eq.~(\ref{eq:BDMPSsimple}), is valid in the multiple scattering regime. It is characterized by the maximal gluon induced energy $\omega_c= \qhat L^2/2$. The spectrum is regulated in the infrared when $t_\text{br}$ is of the order of the mean free path $\lambda_\text{mfp}$, which corresponds to the Bethe-Heitler (BH) frequency $\omega_\text{BH}=\hat q \lambda_\text{mfp}^2 $. We model this regime by regularizing the branching time, i.e., $t_\text{br} \to t_\text{br}+\lambda_\text{mfp}$ and refer to \cite{Blaizot:2013hx,Blaizot:2013vha} for further details on the derivation of Eq.~(\ref{eq:re_1}). 

\begin{figure}
\centering
\includegraphics[width=0.5\textwidth]{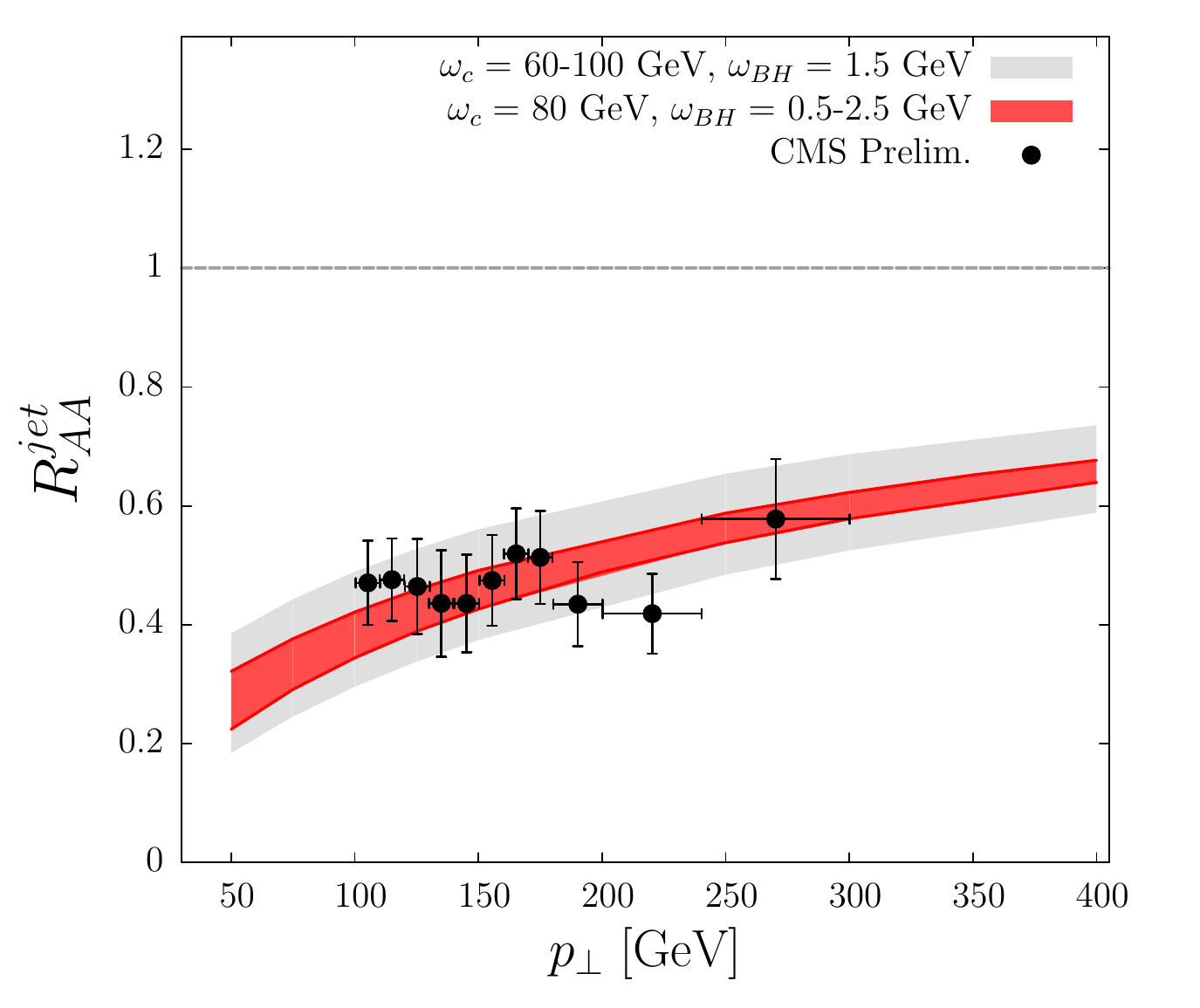}
\caption{Calculation of the quenching factor with $\omega_c$ = 80 GeV, Eq.~(\ref{eq:QuenchingFactor}), as a funtion of jet $p_\perp$ for central Pb-Pb collisions. The dark (red) band includes the variation of $\omega_\text{BH}$ around a central value of 1.5 GeV. The light (grey) band includes, in addition, a variation of $\omega_c \in$ [60, 100] GeV. The experimental data are taken from \cite{CMS:2012rba}.}
\label{fig:RAA}
\end{figure}

The extracted distribution $D^\text{med}_q(x,p_\perp,L)$ of quarks originated from a quark is used to compare the results from Eq.~(\ref{eq:QuenchingFactor}) with experimental data on the nuclear modification factor of fully reconstructed jets in 0--10\% central collisions from CMS \cite{CMS:2012rba}. We have allowed  $\omega_\text{BH}$  to vary between 0.5 and 2.5 GeV to gauge the uncertainty related to the infrared sector. This allows us to extract a value of $\omega_c$ = 80 GeV, see Fig.~\ref{fig:RAA}. For the purpose of illustration, we have also studied the sensitivity to $\omega_c$ by allowing it vary around the central value, see Fig.~\ref{fig:RAA}.

In order to settle on a self-consistent set of parameters, we will from here on use a mean jet path length of $L = $ 2.5 fm for 0--10\% central Pb-Pb collisions. This choice is slightly reduced compared to the typical root mean square of the nuclear overlap in central Pb-Pb collisions motivated by the inherent surface bias of inclusive jet observables \cite{Polosa:2006hb}. The value of $L$ together with the extracted value of $\omega_c$ allows to relate all remaining medium parameters. We notice further that all relevant  parameters vary only mildly within the range of relevant $L$ values and can therefore be expected to be well described by their average values. For example, we extract the average transport coefficient $\qhat =$ 5.1 GeV$^2$/fm. 

A crucial feature of the rate equation Eq.~(\ref{eq:re_1}) is that it describes quasi-democratic branchings of soft gluons and leads to turbulent flow of energy up to large angles \cite{Blaizot:2013hx}. A particularly suited observable to study these effects is therefore the fraction of jet energy still remaining inside a cone defined by the jet reconstruction radius. We calculate this quantity by
\beq
\label{eq:eincone}
\mathcal{E} \left(\theta< \Theta_\text{jet}\right)\equiv \int_0^1 \d x  \, \int_0^{\Theta_\text{jet}} \!\!\d\theta\,\sum_{i=q,g} x\frac{\d N^\text{med}_i }{\d \theta\,\d x}\,,
\eeq
which sums the energy of partons inside the jet cone, i.e.,  $\theta<\Theta_\text{jet}$. In terms of transverse momenta this limitation corresponds to $k_\perp <  xQ$. On the other hand, the typical transverse momentum of a parton propagating in the plasma is given by the characteristic scale $Q_s = \sqrt{\qhat L}$. Hence, the angular condition can be turned into a condition on the parton energy, $x>x_0$, where $x_0 \equiv Q_s/Q$.  Hence, we shall approximate, $\mathcal{E} \left(\theta< \Theta_\text{jet}\right)\approx \mathcal{E}\left( x>x_0\right)$. In our case $Q_s =$ 3.6 GeV severely restricts the amount of soft induced radiation that is allowed within the cone. The description of broadening will be discussed in more detail in a forthcoming work, see also \cite{Blaizot:2013vha}.

We have computed the in-cone energy fraction for two jet reconstruction angles using the previously extracted medium parameters within the uncertainty due to the variation in $\omega_\text{BH}$. We find that up to 14--19\% of the energy flows out a cone of $\Theta_\text{jet} = 0.3$ ($ x_0=0.12$). We scarcely recover more energy by opening the jet cone to $\Theta_\text{jet} = 0.8$ ($x_0=0.045$), in which case roughly 9--15\% of the energy is still missing. This confirms that multiple branching in the medium is an effective mechanism that transports energy from hard to soft quanta at large angles \cite{Blaizot:2013hx}. The results obtained here agree qualitatively with the estimates from the CMS collaboration on the out-of-cone energy flow for di-jets where it was observed that the energy imbalance could be recovered only at angles larger than 0.8 and were carried by tracks with 0.5 GeV $<p_\perp <$ 4 GeV \cite{Chatrchyan:2011sx}. Moreover, the typical transverse momentum broadening of the coherent jet due to scatterings in the plasma is of the order of $Q_s$. Hence, one can estimate the angular deviation of the sub-leading jet to be $\Delta \Theta_\text{jet}\sim Q_s/p_\perp\sim 0.036 $ for a jet $p_\perp=100$ GeV. We note that, $\Delta \Theta_\text{jet}\ll \Theta_\text{jet}$, in agreement with the observation that most di-jets are back-to-back.  

Finally, we focus on the modifications of the fragmentation functions of jets. Concretely we will concentrate on the so called intra-jet energy distribution of hadrons  $\d N^\text{vac}\big/ \d \ln(1/x) \equiv D^\text{vac}(x;Q)$ which is typically plotted in terms of the variable $\ell = \ln (1\big/x_h)$ where $x_h=\sqrt{x^2+(m_h/p_\perp)^2}$ and $x$  are ratios of the hadron and parton energies to the jet energy, respectively. The $Q$ dependence of $D^\text{vac}$ is governed by the Modified Leading-Logarithmic Approximation (MLLA) evolution equations \cite{Dokshitzer:1991wu,*Khoze:1996dn} which take into account the double logarithmic contributions (DLA) as well as the full set of single logarithmic corrections. One of the key features of this evolution is the angular ordering (AO) of subsequent emissions which is a manifestation of color coherence. 
The evolution takes place between the jet scale $Q$ and the hadronization scale $Q_0$ which can be set to $\Lambda_\text{QCD}$ by invoking the Local Parton-Hadron Duality hypothesis.
The resulting parton spectrum can then be directly compared to hadron spectra by introducing an energy independent scaling factor. 

The collimation property of vacuum jets can be inferred directly from the fact that $D^\text{vac}$ only depends on the jet energy and cone angle in terms of $Q$, which is the largest scale of the process. The separation of intrinsic jet and medium scales allow to find the modified fragmentation function directly via the jet calculus rule,
\beq
\label{eq:JetMedCalculus}
D^\text{coh}_\text{med}(x;Q,L) = \int_x^1 \frac{\d z}{z} D^\text{vac}\left(\frac{x}{z};Q \right) D^\text{med}_{q}(z,\pT, L) \,,
\eeq
where $D^\text{med}_{q}(x,p_\perp,L)$ is the distribution of primary quarks \footnote{The multiplicity of gluons originating from medium-induced quark-gluon conversion, and remaining inside the cone is neglected. Their evolution in vacuum is negligible compared to that of the quark since it is limited by the medium scale $Q_s\ll Q$.}.
Here we point out two crucial points concerning Eq.~(\ref{eq:JetMedCalculus}). First and foremost, the subscript of the resulting distribution refers to the coherent jet (color) structure that survives the medium interactions at this level of approximation. In other words, vacuum and medium fragmentation take place independently of each other and are governed by separate evolution equations. Secondly, we have also neglected the variation of the intrinsic jet scale which comes about due to the energy loss at large angles discussed above. As this was estimated to contribute to a $\sim$ 20\% variation to the jet scale, we will allow for a certain variation of the jet energy scale of the medium-modified jets.

\begin{figure}[t!]
\centering
\includegraphics[width=0.5\textwidth]{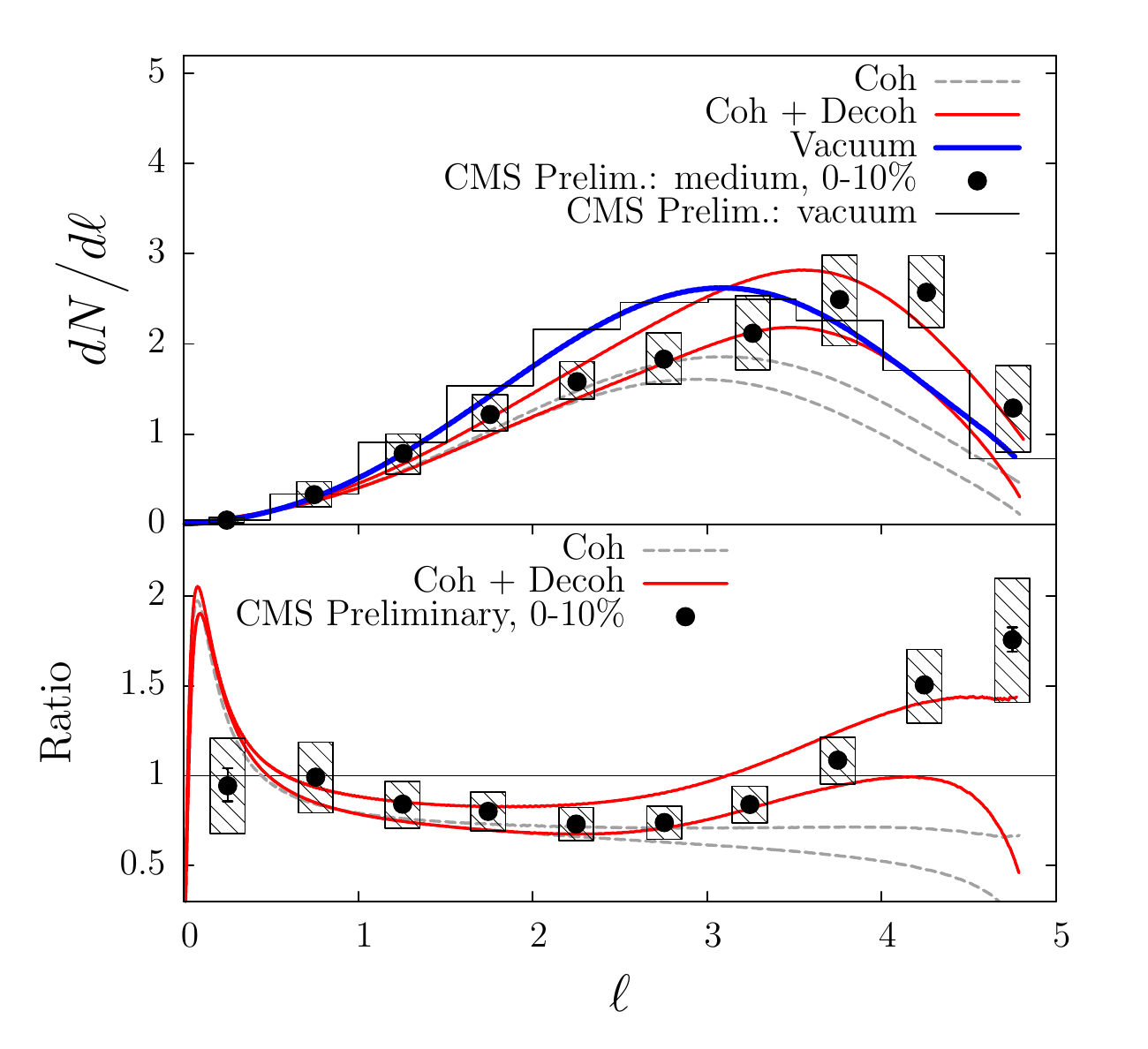}
\caption{Upper panel: the longitudinal fragmentation function plotted as a function of $\ell = \ln 1\big/x$. Lower panel: the ratio of medium-modified and vacuum fragmentation functions. The experimental data are taken from \cite{CMS:2012vba}. See text for further details.
}
\label{fig:FFratio}
\end{figure}

Remarkably, the simple picture incorporated in Eq.~(\ref{eq:JetMedCalculus}), which has shown to be quite consistent up to now, breaks down in the soft sector (cf. grey band in Fig. \ref{fig:FFratio}). This can be traced back to the transverse momentum broadening of soft quanta, Eq.~(\ref{eq:eincone}), which practically removes them from the cone.
However, by comparing the minimal angle for induced radiation $\Theta_c = (\qhat L^3\big/12)^{-\onehalf}$ \cite{Baier:1994bd,Blaizot:2013hx}, which with our set of parameters corresponds to $\sim 0.08$, to the typical jet reconstruction radius, presently considered to be $\Theta_\text{jet} = 0.3$. This implies that sub-leading structures of the jet are resolved by the medium \cite{MehtarTani:2010ma,*MehtarTani:2011tz,CasalderreySolana:2012ef}. Postponing for the moment a more refined treatment of jet energy loss, we will rather emphasize how this breakdown of jet color coherence, initially studied in \cite{CasalderreySolana:2012ef}, demands a more subtle and novel treatment of soft gluon emission at relatively small angles.

Up to now, we have neglected the fact that the jet-medium interactions give rise to additional radiation that violates the strict AO of the vacuum shower \cite{MehtarTani:2010ma,*MehtarTani:2011tz}. Since this component is geometrically separated from the AO vacuum-like radiation and associated with large formation times, it is therefore not affected by the medium (e.g. by transverse momentum broadening). Note that since this contribution also is subleading in DLA, it is enough to include the effect from the first nontrivial splitting. This allows us to add this contribution incoherently to the full, medium-modified intrajet distribution. The intrajet distribution in heavy-ion collisions can thus be written as the sum of two components,
\beq
\label{eq:JetMedTot}
D^\text{jet}_\text{med}(x;Q,L) = D^\text{coh}_\text{med}(x;Q,L) + \Delta D^\text{decoh}_\text{med}(x;Q, L)\,,
\eeq
where $D^\text{coh}_\text{med}$ is the coherent modified jet spectrum found from Eq.~(\ref{eq:JetMedCalculus}) and the decoherence of in-cone vacuum radiation is contained in $\Delta D^\text{decoh}_\text{med}$. We compute the real contribution at two successive emissions at DLA accuracy with the inclusion of running coupling effects, yielding
\begin{multline}
\label{eq:GAAO_1}
\Delta D^\text{decoh}_\text{med}(x;Q,\hat q, L) =  \int_\omega^E \frac{\d \omega'}{\omega'} 
 \int_{Q_0/\omega}^{\Theta_\text{jet}}  \frac{\d \theta'}{\theta'} \\ \times \Delta_\text{med}(\theta')\,\alpha_s(\omega'\theta') 
 \int_{\theta'}^{\theta_\text{max}} \,\frac{\d \theta}{\theta} \, \alpha_s(\omega\theta)\,,
\end{multline}
where the decoherence parameter reads $\Delta_\text{med} (\theta')= 1- \exp[-\theta' \qhat L^3/12]$ \cite{MehtarTani:2010ma,*MehtarTani:2011tz}  and $\theta_\text{max}=\min(\Theta_\text{jet}, Q_\text{med}/\omega)$. In this context, $Q_\text{med} = \max\left[(\theta' L)^{-1},Q_s \right]$ is the hardest scale of the splitting. To test the sensitivity of the resulting distribution to this parameter we have varied $Q_\text{med}$ while keeping $Q_s$ at the central value such that $0.8 < \ln Q_\text{med}\big/ Q_0 < 3.2$. As a further refinement, we will also demand that the first splitting occurs inside the medium. This puts a constraint on the formation time of the first gluon, i.e., $t_\text{f}(\omega') \simeq (\omega' \theta'^2)^{-1} < L$. For consistency, we will also count the traversed path length from the production point by shifting $L \to L - t_\text{f}(\omega')$ in $\Delta_\text{med}(\theta')$.

The resulting vacuum and medium distributions for jets with $Q=$ 30 GeV are shown in the upper panel of Fig. \ref{fig:FFratio}, while the lower panel details the ratio of the latter to the former. We compare to experimental data from CMS for jets with $\pT > $ 100 GeV \cite{CMS:2012vba}. First, the vacuum baseline data are reproduced by the MLLA equation by adjusting the relevant parameters ($Q_0 =$ 0.4 GeV, $m_h = $ 1.1 GeV and $K =$ 1.6) to optimize the description, depicted by a sold (blue) line in the upper part of Fig.~\ref{fig:FFratio}. Due to the energy loss in the medium, we have allowed the jet scale of the medium-modified jets to vary within $E \in$ [100,125] GeV (we plot the results for the extreme cases). In what follows, the variation of the BH frequency was found to be negligible and the central value $\omega_\text{BH} =$ 1.5 GeV was used. The result of using only Eq. (\ref{eq:JetMedCalculus}), depicted by the dashed (grey) lines, which assume coherent radiation, yields a suppression of the distribution at all $\ell$ as compared to that in vacuum. This reflects the energy loss via soft gluon radiation at large angles off the total charge of the jet and is in agreement with the suppression of the nuclear modification factor. However, the data indicates that the suppression turns into an enhancement when $\ell \gtrsim 3$ in the most central collisions \cite{CMS:2012vba}. Accounting for color decoherence as given in Eq.~(\ref{eq:JetMedTot}) we describe the excess of soft particles in the measured medium-modified fragmentation function, see the thin-solid (red) curves in Fig.~\ref{fig:FFratio}. The resulting ratio of medium-to-vacuum distributions show the characteristic dip and enhancement behavior with increasing $\ell$ around the humpbacked plateau. Note that the MLLA equation is valid at intermediate values of $\ell$ and that the region of small $\ell\lesssim 1$ is sensitive to energy conservation and hence should be discarded. On the other hand, for $\ell\sim 4.5$ the distribution in reaching the limits of phase space and is very sensitive to non-perturbative physics and the precise jet energy scale.

To summarize, we have investigated several jet observables that have recently been measured at the LHC. Our model based on the QCD limit of color coherence is consistent with the different features seen in data and we are able to pin down departures from this picture in the soft sector of fragmentation functions, which we argue is an evidence for partial decoherence. Our approach further shows how jets produced in these collisions can be used as a powerful tool to extract information about the QGP and color coherence.

\section*{Acknowledgments}
K. T. is supported by a Juan de la Cierva fellowship and by the research grants FPA2010-20807, 2009SGR502 and by the Consolider CPAN project. Y. M. -T. is supported by the European Research Council under the Advanced Investigator Grant ERC-AD-267258.


\end{document}